\documentstyle[preprint,aps]{revtex}
\tighten
\draft
\begin{document}
\widetext
\input epsf
\preprint{CLNS 97/1468, HUTP-97/A016, NUB 3138}
\bigskip
\bigskip
\title{Phenomenology of $3$-Family Grand Unified String Models}
\medskip
\author{Zurab Kakushadze$^{1,2}$\footnote{E-mail: 
zurab@string.harvard.edu}, 
Gary Shiu$^3$\footnote{E-mail: shiu@mail.lns.cornell.edu}, S.-H. Henry Tye$^3$\footnote{E-mail: tye@mail.lns.cornell.edu}, 
Yan Vtorov-Karevsky$^3$\footnote{E-mail: gilburd@mail.lns.cornell.edu}}

\bigskip

\address{$^1$Lyman Laboratory of Physics, Harvard University, Cambridge, 
MA 02138\\
$^2$Department of Physics, Northeastern University, Boston, MA 02115\\
$^3$Newman Laboratory of Nuclear Studies, Cornell University,
Ithaca, NY 14853-5001}
\date{\today}
\bigskip
\medskip
\maketitle

\begin{abstract}
{}In the $3$-family grand unified string models constructed so far,
there is only one adjoint (and no higher dimensional representation) Higgs 
field in 
the grand unified gauge group. In this preliminary analysis, we address
the proton-decay problem in the $3$-family $E_6$ and related $SO(10)$ 
string models. 
In particular, we analyze the doublet-triplet splitting (within certain 
assumptions about
non-perturbative dynamics). It appears that generically some fine-tuning 
is necessary to arrange for a pair of Higgs doublets to be light, while 
having color Higgs triplets superheavy. We also discuss charge-$2/3$ 
quark mass matrix that generically also seems to require some 
fine-tuning to have rank $1$.
 
\end{abstract}
\narrowtext

{}One elegant way for superstring theory to include the standard model 
in its low energy effective theory \cite{revw} is via extensively
studied supersymmetric (SUSY) grand unified theory (GUT) \cite{review}.
Spontaneous symmetry breaking of the grand unified gauge group to the
standard model in the low energy effective field theory requires at least 
an adjoint (or other appropriate higher dimensional representation) Higgs field in the light
mass spectrum. 
This requirement imposes a strong constraint on GUT string model-building.
In addition, dynamical SUSY breaking via gaugino condensation requires an 
asymptotically-free hidden sector (SUSY breaking is then transmitted to 
the observable sector via interactions with gravity or other
messenger/intermediate sector \cite{gaugino}).
Recently, two of us have constructed 
grand unified string models that satisfy the above requirements and have 
three chiral families of fermions in the GUT gauge group \cite{kt}. 
In fact, a classification of such $3$-family grand unified models was
carried out within the framework of conformal (free) field theory and 
asymmetric orbifolds. The form of the superpotentials for these models 
was determined recently \cite{kst}. In this letter, we give a
preliminary analysis of the phenomenological properties of some of 
these models. More specifically, we 
analyze the proton decay
problem and the fermion mass matrix in 
the $3$-family $E_6$ and the related $SO(10)$ string models. 

{}The $E_6$ and $SO(10)$ GUTs \cite{gut} are well studied. 
Here, the string models are very specific realizations of these GUTs.
Within the string model-building framework described above, we find only 
one $3$-family $E_6$ model. In this paper, we make a preliminary study 
of the phenomenological
properties of this model and the closely related $SO(10)$ string model.  
They have 5 right-moving and 2 left-moving fermion families.
Upon generation--antigeneration pairings, one is left with three chiral
families. Analysis shows that one pairing takes place at around 
the $M_{GUT}$ scale, while the other pairing occurs at a (perhaps, much) 
lower scale.

The problem of color Higgs mediated proton decay \cite{wein} has a simple 
solution in the framework of $SO(10)$ models. This is achieved by 
giving appropriate {\em vev}
to the adjoint (for example, via the Dimopoulos--Wilczek mechanism 
\cite{dim}). 
In the model we considered, the Higgs doublets and triplets 
generically become massive at the 
scale slightly below $M_{GUT}$ 
(with the exception of a pair of doublets
and a pair of triplets which may become massive at a much lower scale; the 
latter triplets are harmless for proton stability, however).
As the string models have only one adjoint Higgs field, 
one can arrange for a pair of Higgs doublets to be light, while having 
color Higgs triplets
heavy. This, however, requires certain fine-tuning since under spontaneous
symmetry breaking of $E_6$ to the standard model, all Higgs doublets and
triplets are heavy in a generic point in the moduli space. To obtain one
light Higgs doublet, the moduli have to satisfy one constraint.
In terms of the moduli entering the Higgs doublet mass matrix 
(which is deduced from the approximate superpotential used in this paper), 
this means a light Higgs
doublet exists in a 10-dimensional sub-manifold in the original
11-dimensional moduli space. This is fine-tuning.

{}We should emphasize that our conclusions rely 
heavily on important assumptions about the non-perturbative 
string dynamics, an issue we shall only comment on briefly.
Furthermore, some mixings of fields are ignored in this preliminary analysis.

{}To begin, let us look at the massless spectra of the $E_6$ and 
$SO(10)$ models \cite{kt} given in the following table.
The gravity, dilaton and gauge supermultiplets are not shown.
The $E_6$ model has gauge symmetry 
$SU(2)_1 \otimes (E_6)_3\otimes U(1)^3$, while the $SO(10)$ model has
gauge symmetry $SU(2)_1 \otimes SO(10)_3\otimes U(1)^4$. 
The subscripts label the levels of the current algebra realizations of the
gauge symmetries. In practice, this means that the fine structure constant 
of the level-$3$ GUT gauge symmetry is $1/3$ that of the level-$1$ 
$SU(2)$ gauge symmetry. The $U(1)$ charge of a particle is 
given by its integer value in the table multiplied by the normalization 
provided at the bottom of the table. 
Note that all massless particles with 
$E_6$ quantum numbers are singlets under $SU(2)$, {\em i.e.}, this 
$SU(2)$ plays the role of a hidden sector. The $U(1)$s play the role of 
intermediate/messenger symmetries. 
\begin{center}
\label{models}
\begin{tabular}{|c|c||c|c|}
\hline
  & & & \\
 $E1$ & & $T1(1,1)$ &\\
 $SU(2) \otimes E_6 \otimes U(1)^3$ & Field 
   & $SU(2) \otimes SO(10) \otimes U(1)^4$ & Field
      \\ \hline
  & & &\\
    $ ({\bf 1},{\bf 78})(0,0,0)_L$ & ${\widehat\Phi}$
 &  $ ({\bf 1},{\bf 45})(0,0,0,0)_L$ & $\Phi$ \\
&&$ ({\bf 1},{\bf 1})(0,0,0,0)_L$ & $\phi$ \\
 $ ({\bf 1},{\bf 1})(0,+6,0)_L$ & $U_{0}$ 
   & $({\bf 1},{\bf 1})(0,+6,0,0)_L$ & $U_{0}$\\
    $ \hspace{-1mm} 2 ({\bf 1},{\bf 1})(0,-{3},\pm 3)_L$ 
   & $U_{+ \pm}, U_{- \pm}$  
   & $ \hspace{-1mm} 2 ({\bf 1},{\bf 1})(0,-{3},\pm 3,0)_L$
   & $U_{+ \pm}, U_{- \pm}$ 

\\
   & & &\\
  \hline
  & & & \\
     $ ({\bf 1},{\bf 27})(0,-{2},0)_L$ & $\chi_{0}$ 
   &  $ ({\bf 1},{\bf  16})(0,-{2},0,-1)_L$ 
   & ${\bf Q}_0$\\
    & &$ ({\bf 1},{\bf  10})(0,-{2},0,+2)_L$ & ${\bf H}_0$ \\
 & &$ ({\bf 1},{\bf  1})(0,-{2},0,-4)_L$ & ${\bf S}_0$ \\
    $\hspace{-1mm} 2({\bf  1}, {\bf 27})(0,+1,\pm1)_L$ 
   & $\chi_{+ \pm}, \chi_{- \pm}$ 
   & $\hspace{-2mm} 2({\bf  1}, {\bf  16})(0,+1,\pm1,-1)_L$ 
   & ${\bf Q}_{+\pm},{\bf Q}_{-\pm}$\\
    & &$\hspace{-2mm} 2({\bf  1}, {\bf 10})(0,+1,\pm1,+2)_L$ 
   & ${\bf H}_{+\pm},{\bf H}_{-\pm}$  \\
    & &$\hspace{-4mm} 2({\bf  1}, {\bf 1})(0,+1,\pm1,-4)_L$ 
   & ${\bf S}_{+\pm},{\bf S}_{-\pm}$ \\
  & & & \\
 \hline
   & & & \\
    $ ({\bf 1},{\overline {\bf 27}}) (\pm 1,-1,0)_L$ 
      &  $ \widetilde{\chi}_{\pm}$
      &  $ ({\bf 1},{\overline {\bf 16}}) (\pm 1,-1,0,+1)_L$ 
      &  ${\bf\widetilde Q}_\pm$\\
       & & $ ({\bf 1},{{\bf 10}}) (\pm 1,-1,0,-2)_L$ 
      &  ${\bf\widetilde H}_\pm$  \\
       & & $ ({\bf 1},{{\bf 1}}) (\pm 1,-1,0,+4)_L$ 
      &  ${\bf\widetilde S}_\pm$  \\
     & & & \\
 \hline
  & & & \\
    $({\bf 2},{\bf 1})(0,0,{\pm 3})_L$ & $D_{\pm}$
        & $({\bf 2},{\bf 1})(0,0,{\pm 3},0)_L$ & $D_{\pm}$ \\
         $ ({\bf 1},{\bf 1})(\pm {3},+{3},0)_L$  
        & $\widetilde{U}_{\pm}$ 
        & $ ({\bf 1},{\bf 1})(\pm {3},+{3},0,0)_L$  
        & $\widetilde{U}_{\pm}$ \\
   & & & \\
   \hline
  & & & \\
  $r_{U(1)}:~(1/ \sqrt{6}, ~1/{3\sqrt{2}}, ~1/\sqrt{6})$ & 
        & $(1/ \sqrt{6}, ~1/{3\sqrt{2}},  ~1/\sqrt{6},~1/6)$ &\\
  & & & \\
\hline
\end{tabular}
\end{center}

\bigskip
\bigskip
 
{}The $E_6$ model has $5$ left-handed ${\bf 27}$s: 
$\chi_{0}$, $\chi_{++}$, $\chi_{--}$, $\chi_{+-}$ and $\chi_{-+}$, and 
$2$ left-handed ${\overline {\bf 27}}$s: $\widetilde{\chi}_{+}$ and 
$\widetilde{\chi}_{-}$. Their lowest order non-vanishing couplings in 
the superpotential are \cite{kst}
 \begin{eqnarray}
\label{e6-super} 
W&=&\lambda_0\;U_0
               \left( U_{++}U_{--} + U_{+-}U_{-+}\right)
    +\lambda_1\;\chi_{0} \left(\chi_{++}\chi_{--} +
 \chi_{+-}\chi_{-+}\right){\widehat\Phi} \nonumber \\[2mm]
 &+& \lambda_2\;\widetilde{\chi}_{+}\widetilde{\chi}_{-}
\left(\chi_{++}\chi_{--} + \chi_{+-}\chi_{-+}\right) 
 + \lambda_3\left(\chi_{++}^3U_{--} + \chi_{+-}^3U_{-+} + 
\chi_{-+}^3U_{+-} + \chi_{--}^3U_{++}\right){\widehat\Phi}^2\nonumber\\[2mm]
&+&\lambda_4\;\chi_0^3\,U_0D_+D_-{\widehat\Phi}^2 +
 \lambda_5\left(\widetilde{\chi}_{+}^3{\widetilde U}_- +
\widetilde{\chi}_{-}^3{\widetilde U}_+\right)D_+D_-{\widehat\Phi}\,+\,\ldots
\end{eqnarray}
where the traces over the irreps of the gauge group are implicit here. 
The $\lambda_{k}$ are polynomials of ${\widehat\Phi}^3$, with 
$\lambda_{k} [0]\neq 0$. 
In principle, one can extract the $\lambda_{k}[{\widehat\Phi}^3]$ polinomials
from tree level string scatterings; they may also be determined using 
string modular invariant properties in a much simpler way.
If all the ${\em vev}$s
are below the string scale it is a reasonable approximation to ignore the
higher order terms in the superpotential.
It is clear that the gauge discrete symmetries in superstring theory
impose powerful selection rules on the couplings. {\em A priori}, the
superpotential may contain other gauge invariant terms such as the
three-point couplings $\chi_{0} \chi_{\alpha +} \chi_{\beta -}$ where
$\alpha, \beta= \pm$. However, these terms are forbidden by the
stringy discrete symmetries of the model \cite{kst}. Similarly,
couplings such as $\chi_{0}\chi_{++}\chi_{+-} {\widehat\Phi}$ are
not allowed even though they contain gauge singlets.

{}The spectra of the $SO(10)$ and $E_6$ models
are very similar. In particular, the spectrum of the 
$SO(10)$ model is the 
same as that of the $E_6$ model with a non-zero {\em vev} of the 
adjoint ${\widehat\Phi}$ of $(E_6)_3$ such that $(E_{6})_3$ is broken down to
$SO(10)_3 \otimes U(1)$ (the last $U(1)$ in the $SO(10)$ part 
of the table). Thus, the ${\bf 27}$ of $(E_6)_3$ branches into 
${\bf 16}(-1)+{\bf 10}(+2)+ {\bf 1}(-4)$ of $SO(10)_3 \otimes U(1)$ 
(denoted as ${\bf Q,\;H}$ and ${\bf S}$ respectively). 
The adjoint ${\bf 78}$ branches into 
${\bf 45}(0)+{\bf 1}(0)+{\bf 16}(+3)+{\overline {\bf 16}}(-3)$. 
Note that while the ${\bf 45}(0)$ and ${\bf 1}(0)$ are present in the $SO(10)$ 
model, the ${\bf 16}(+3)$ and ${\overline {\bf 16}}(-3)$ are missing because
of the super-Higgs mechanism. 

{}From the above discussion, it is clear that to get the superpotential for 
the $SO(10)$ model 
from that of the $E_6$ model, we can simply replace
$\chi$s by $({\bf Q+H+S})$s (and similarly 
for ${\widetilde \chi}$s), and ${\widehat\Phi}$ 
by $\Phi +\phi$ in (\ref{e6-super}). 
The allowed 4-point couplings in the $SO(10)$ superpotential are:
\begin{eqnarray} \label{sp-so10} 
 W &=& \lambda_{11}\left\{{\bf S}_{0}({\bf H\cdot H}) 
+ {\bf H}_{0}({\bf S\cdot H})\right\}(\Phi + \phi )
 + \lambda_{12}\left\{{\bf H}_{0}({\bf Q\cdot Q}) 
+ {\bf Q}_{0}({\bf H\cdot Q}) 
\right\}(\Phi + \phi ) \\[3mm] 
&+&  \lambda_{21}\,\widetilde{\bf S}_{+}\widetilde{\bf S}_{-} ({\bf S\cdot S})
   + \lambda_{22}\,\widetilde{\bf Q}_{+}\widetilde{\bf Q}_{-}({\bf Q\cdot Q})
+ \lambda_{23}\left(\widetilde{\bf S}_{+}\widetilde{\bf Q}_{-} 
 + \widetilde{\bf S}_{-}\widetilde{\bf Q}_{+}\right)({\bf S\cdot Q})
+\nonumber\\[3mm]
&+& \lambda_{24}\left(\widetilde{\bf S}_{+}\widetilde{\bf H}_{-} + 
\widetilde{\bf S}_{-}\widetilde{\bf H}_{+}\right)({\bf S\cdot H})
+ \lambda_{25}\left(\widetilde{\bf H}_{+}\widetilde{\bf Q}_{-} 
 + \widetilde{\bf H}_{-}\widetilde{\bf Q}_{+}\right)({\bf H\cdot Q})
\nonumber\\[3mm] 
&+&\lambda_{26}\,\widetilde{\bf H}_{+}\widetilde{\bf H}_{-}({\bf H\cdot H})
+ \lambda_{27}\left\{\widetilde{\bf Q}_{+}\widetilde{\bf Q}_{-}({\bf S\cdot H})
+ (\widetilde{\bf S}_{+}\widetilde{\bf H}_{-} + 
\widetilde{\bf S}_{-}\widetilde{\bf H}_{+})({\bf Q\cdot Q}) \right\}+ \ldots
\nonumber\end{eqnarray}
Here, we introduce the following notation:
\begin{eqnarray}
({\bf A\cdot B}) \equiv  
{\bf A_{++}B_{--} + B_{++}A_{--} + A_{+-}B_{-+} + B_{+-}A_{-+}}
\end{eqnarray}
where $A$ and $B$ could be any of the fields ${\bf S, H, Q\,}$. 
Only the terms originating from the $\lambda_1$- and $\lambda_2$--terms in the
$E_6$ superpotential (\ref{e6-super}) are shown. The couplings 
$\lambda_{1i}$ stem from the $\lambda_1$--term in the $E_6$ superpotential 
and are {\em a priori} different because of the
Clebsch--Gordan coefficients which arise upon breaking $E_6$ to 
$SO(10) \otimes U(1)$.
The same is true for the $\lambda_{2i}$ couplings. 

{}Classically, the adjoint Higgs $\Phi$ is a flat modulus, so it can 
develop an arbitrary {\em vev}. However, it is likely that gaugino condensate 
will lift this flatness. Let us consider the following rather standard 
scenario \cite{gaugino}.
With only one pair of chiral doublets $D_\pm$, the gauge group $SU(2)$ is
asymptotically-free. Assuming that $\alpha_{GUT} \approx 1/24$, we have
$\alpha_{st} = \alpha_{SU(2)} = 3 \,\alpha_{GUT} \approx 1/8$. 
(This factor of $3$ difference 
reflects the level--3 current algebra realization of the GUT symmetry).
In string theory, the coupling $\alpha_{st}$ is 
not a free parameter, but
is determined by the expectation value of the dilaton.
Gaugino condensation in the $SU(2)$ hidden sector generates
an effective scalar potential which, hopefully, stabilizes the dilaton
expectation value in a self-consistent way. (Since globally $N=1$ 
supersymmetric $SU(2)$ with one flavor does not have a quantum vacuum, 
one might hope that including non-perturbative corrections to the
K\"{a}hler potential \cite{bd} would stabilize 
the moduli and the dilaton).
Here we shall simply make this assumption and consider 
the consequences. The scale of gaugino condensation is given by
\begin{eqnarray}
\mu^2 = M^2_{st} \exp\left({\frac{-4\pi}{b_0 \alpha_{st}}}\right) \approx
\left(10^{13}GeV\right)^2 
\end{eqnarray}
where $b_0 = 6-1 =5$ and the string scale $M_{st}\approx 10^{18} ~GeV$.
This yields the supersymmetry breaking scale in the visible sector 
\cite{nath}
\begin{eqnarray}
m_S \approx \frac{\mu^3}{M^2_{st}} \approx ( 10\sim 100 )~TeV
\end{eqnarray}
It is worth noticing that $m_{S}$ is very sensitive to $\alpha_{st}$.
The above estimate is tightly related to the fact that the GUT symmetry
is realized at level $3$.

{}Due to the presence of the K\"{a}hler potential $K(z, \bar{z})$ in the
$F$-term of the effective scalar potential $V_F$:
\begin{eqnarray}
V_F(z, \bar{z}) = e^{K/M^2_{pl}}\left( D_z W K^{-1}_{z\bar{z}}
\overline{D_z W} - 3 \frac{|W|^2}{M_{pl}^2}\right)
\end{eqnarray}
(where $D_z W \equiv W_z + WK_z/M_{pl}^2$, with $M_{pl}$ being the Planck 
mass) 
the gaugino condensate will
generate an effective potential for the adjoint Higgs field. 
The 
height of this potential term is rather small, of the order of $m_S^4$. 
Dimensional arguments suggest that the $SO(10)$ adjoint Higgs may develop 
string (or GUT) scale {\em vev}; at the same time, 
the adjoint Higgs fields of the remaining gauge symmetry (such as QCD) 
will have masses of the order $m_S$. These relatively light Higgs fields
are a generic prediction of string GUT models. (There may exist 
model-dependent mechanisms for generating intermediate masses for the 
adjoints, but we will not address this issue here). 
Because of the difficulties in the analysis of dynamical symmetry breaking,
we shall simply treat the adjoint Higgs as flat moduli.

{}Let us consider the possible scenarios for breaking the 
$SO(10)$ down to that of the standard model, 
{\em i.e.}, $SU(3)_c \otimes SU(2)_w \otimes U(1)_Y$. 
There is one adjoint Higgs field in the massless spectrum of 
the model. The adjoint must break $SO(10)$ down to its subgroups that 
contain $SU(3)_c \otimes SU(2)_w \otimes U(1)_Y$. Here we are interested in
the breaking $SO(10)\supset SU(3)_c \otimes SU(2)_w \otimes 
U(1)_Y \otimes U(1)$ by turning on the {\em vev} of the adjoint
$\Phi =\epsilon\otimes {\mbox{diag}}(a,a,a,b,b)$, where 
$\epsilon$ is a $2\times 2$ antisymmetric matrix and $a,b\not=0$, $a\not=b$.

{}The adjoint Higgs field of $SO(10)$ does not carry any other gauge 
quantum numbers, 
so there is no danger of mixing the $SO(10)$ quantum numbers with other 
gauge quantum numbers in the adjoint breaking.
Since the adjoint by itself cannot break the $SO(10)$ gauge 
symmetry down to that of the standard model, some other 
fields must acquire {\em vev}\,s to achieve further breaking of the 
gauge symmetry. 
These fields must carry $SO(10)$ quantum numbers, and contain 
 $SU(3)_c \otimes SU(2)_w$ singlets, {\em i.e.}, they must come from 
${\bf 16}+{\overline {\bf 16}}$ of $SO(10)$.

{}Let us denote the values of the {\em vev}s of ${\bf S}_0, 
{\bf S}_i, {\bf Q}_0, {\bf Q}_i, 
{\bf\widetilde S}_\pm$ and ${\bf\widetilde Q}_\pm$ as $s_0, s_i, q_0, q_i, 
{\widetilde s}_\pm$ and ${\widetilde q}_\pm$ correspondingly. From 
the $D$-term of the effective scalar potential corresponding
to a $U(1)$ gauge group:
\begin{equation}
V_{D}= {\alpha_{st}\pi} (\sum_{A} X_A \langle \phi_A \rangle^2 )^2
\end{equation}
(where $X_A$ is the charge of field $\phi_A$),
the D-flatness condition follows: 
\begin{equation}
\label{D-flat}
 q_0^2 + q^2 = 
{\widetilde q}^2~,~~~~~{\rm where} 
~~q\equiv \sqrt{\sum_i q_i^2}~,~~~~~ 
{\widetilde q}\equiv \sqrt{{\widetilde q}_+^2 
+ {\widetilde q}_-^2}~,
\end{equation}
This implies at least one of the ${\bf\widetilde Q}_{\pm}$ {\em vev}\,s 
must be non-zero. 
Thus, the ${\bf 16}+{\overline {\bf 16}}$ breaking scale is given by 
${\widetilde q}$.

{}To analyze the $SU(3)_3 \otimes SU(2)_3 \otimes U(1)_Y \otimes U(1)$ spectrum
of the model, let us
consider the following branchings of  the $SO(10)$ irreps 
${\bf  45}$, ${\bf 16}$ and ${\bf 10}$ under the 
breaking $SO(10)_3 \supset SU(3)_3 \otimes SU(2)_3 \otimes U(1)_{Y} \otimes 
U(1)$:
\begin{eqnarray}
{\bf 45}=&&({\bf 1},{\bf 1})(0,0)+[({\bf 1},{\bf 1})(0,0)+({\bf 8},
{\bf 1})(0,0) +({\bf 1}, {\bf 3})(0,0) +
 ({\bf 3},{\bf 2})(-5,0) +({\overline {\bf 3}},{\bf 2})(+5,0)]\nonumber\\
 &+&[({\bf 1},{\bf 1})(+6,+4)+({\overline {\bf 3}},{\bf 1})(-4,+4) +({\bf 3}
, {\bf 2})(+1,+4)]\nonumber\\
\label{45-break}
 &+&[({\bf 1},{\bf 1})(-6,-4)+({\bf 3},{\bf 1})(+4,-4) +({\overline {\bf 3}}
, {\bf 2})(-1,-4)]~,\\
 \label{16-break}{\bf 16}=&&
\stackrel{\textstyle q}{({\bf 1},{\bf 1})(0,-5)} +[
\stackrel{\textstyle D}{({\overline {\bf 3}},{\bf 1})(+2,+3)} 
+\stackrel{\textstyle L}{({\bf 1}, {\bf 2})(-3,+3)}] \\
 &+&[\stackrel{\textstyle E}{({\bf 1},{\bf 1})(+6,-1)} +
\stackrel{\textstyle U}{({\overline {\bf 3}},{\bf 1})(-4,-1)} +
\stackrel{\textstyle Q}{({\bf 3}, {\bf 2})(+1,-1)}]~,\nonumber\\
\label{10-break}
 {\bf 10} =&&[\stackrel{\textstyle H}{({\bf 3},{\bf 1})(-2,+2)}
 +\stackrel{\textstyle h}{({\bf 1}, {\bf 2})(+3,+2)}] +
 [\stackrel{\textstyle H'}{({\overline {\bf 3}},{\bf 1})(+2,-2)} +
\stackrel{\textstyle h'}{({\bf 1}, {\bf 2})(-3,-2)}]~.
\end{eqnarray}
and the corresponding fields from $\tilde{\chi}$ are denoted with a
tilde.

The spontaneous symmetry breaking by the adjoint Higgs field discussed above 
correspond to giving appropriate 
{\em vev}\,s to the two neutral singlets in ${\bf 45}$. 
The ${\bf 16}$ acquires {\em vev} in the direction of  
the singlet $({\bf 1}, {\bf 1})(0,-5)$, leaving $U(1)_Y$ unbroken.
Standard notation is introduced for the quark ($D, U, Q$), 
lepton ($L, E$), Higgs doublet ($h, h'$)
 and color Higgs triplet ($H, H'$) superfields.
Let's rewrite the superpotential (\ref{sp-so10}) in terms of  
these superfields
and $\phi, \Phi, s_i, {\widetilde s}_\pm , q_i$ and ${\widetilde q}_\pm$ 
({\em vev}\,s of $SU(3)_c \otimes SU(2)_w \otimes U(1)_Y$ singlets). 
Generation indices and explicit coupling constants are suppressed for brevity:
\begin{eqnarray} \label{sp-sm} 
W &=& \left\{(\phi\,+\,\Phi)h\,+\,{\widetilde s}{\widetilde h}\,+\, 
{\widetilde q}{\widetilde L}\right\}\,Q\,U\,+\,
\left\{(\phi\,+\,\Phi)h' + {\widetilde s}{\widetilde h}'\right\}
(Q\,D\,+\,E\,L)\\
&+&(\phi\,+\,\Phi)(q\,h\,L\,+\,s\,h\,h')\;
+\;({\widetilde s}\,s\,+\,{\widetilde q}\,q)\left({\widetilde h}\,h'
\,+\,{\widetilde h}'\,h\,+\,{\widetilde L}\,L\right)\,
+\,s{\widetilde q}h'{\widetilde L}\,+\,{\widetilde s}q{\widetilde h}L
\nonumber\\
&+&(\phi\,+\,\Phi)(q\,HD\,+\,s\,HH')\;
+\;({\widetilde s}\,s\,+\,{\widetilde q}\,q)\left({\widetilde H}H'\,+\,
{\widetilde H}'H\,+\,{\widetilde D}\,D\right)\,+
\,s{\widetilde q}H'{\widetilde D}\,+\,{\widetilde s}q{\widetilde H}D
\nonumber\\
&+&\left\{(\phi\,+\,\Phi)H\,+\,{\widetilde s}{\widetilde H}\,+\,
{\widetilde q}{\widetilde D}\right\}(Q\,Q\,+\,E\,U)\;+
\left\{(\phi\,+\,\Phi)H'\,+\,{\widetilde s}{\widetilde H}'\right\} 
(Q\,L\,+\,U\,D)\;+\ldots
\nonumber
\end{eqnarray}

Only the terms relevant for the chiral fermion mass matrices and Higgs
doublet and triplet mass matrices are shown. In particular, the
first line of
(\ref{sp-sm})
shows terms contributing to the top, bottom and lepton mass matrices, 
terms in the second line are responsible for Higgs doublet masses, 
terms in the next line give masses to color 
Higgs triplets. The last line contains terms that are responsible
for unacceptably rapid proton decay if the color Higgs triplets are not
sufficiently massive.
As can be seen from the first line of (\ref{sp-sm}), mixing
of $h$ with ${\widetilde L}$, 
$h'$ with $L$, $H$ with ${\widetilde D}$ and $H'$ with $D$  
affects charge-2/3 quark mass matrix differently than lepton and 
charge-\,--1/3 quark mass matrices. The last two matrices are affected by 
mixing in an identical way. So the leptons and the
charge-\,--1/3 quarks are expected to have the same mass 
matrix at the GUT scale. It is worth noticing that, due to the underlying 
gauge and discrete
symmetries of the string model, the lepton-number-violating terms 
$LLE$ and $LQD$ and the baryon-number-violating term $UDD$ are absent. 

{}We also need to impose the F-flatness 
conditions with respect to the fields ${\bf S}_0,\,{\bf S}_i,\,
{\bf\widetilde S}_{\pm}$, 
${\bf Q}_0,\,{\bf Q}_i$ and ${\bf\widetilde Q}_{\pm}$ 
(note that
the F-flatness conditions with respect to the Higgs fields ${\bf H}_0$, 
${\bf H}_i$ 
and ${\bf\widetilde H}_{\pm}$ are automatically satisfied). 
These conditions can be easily read from 
(\ref{sp-so10}):
\begin{eqnarray}
\label{f1}
 \partial W/ \partial {\bf\widetilde Q}_\mp = 0 &~~~~~\Rightarrow ~~~~~& 
\lambda_{22}\,{\widetilde q}_\pm (q\cdot q) +
 \lambda_{23}\,{\widetilde s}_\pm (s\cdot q) = 0,\\
\label{f2}
\partial W/ \partial {\bf\widetilde S}_\mp = 
0 &\Rightarrow & \lambda_{23}\,{\widetilde q}_\pm 
(s\cdot q) + \lambda_{21}\,{\widetilde s}_\pm (s\cdot s) = 0 ,\\
\label{f3}
 \partial W/ \partial {\bf Q}_{\bar i} = 0 &\Rightarrow & 2\lambda_{22}\,
q_i {\widetilde q}_+ 
{\widetilde q}_- + 
\lambda_{23}\,s_i\left({\widetilde s}_+ {\widetilde q}_- 
+ {\widetilde s}_- {\widetilde q}_+\right) = 0 ,\\
\label{f4}
 \partial W/ \partial {\bf S}_{\bar i} = 0 &\Rightarrow & 
\lambda_{23}\,q_i 
\left({\widetilde s}_+ {\widetilde q}_- 
+ {\widetilde s}_- {\widetilde q}_+\right) 
+ 2\lambda_{21}\,s_i {\widetilde s}_+ {\widetilde s}_- = 0
\end{eqnarray}
Two constraints need to be satisfied in order for a non-trivial solution for 
the F-flatness conditions to exist:

\begin{eqnarray}
\label{flat-1}
0&=&\lambda_{21}\lambda_{22}\,(s\cdot s)(q\cdot q) -
\lambda_{23}^2\,(s\cdot q)^2~,\\
\label{flat-2}
0&=&4\,\lambda_{21}\lambda_{22}\,{\widetilde s}_+{\widetilde s}_-
{\widetilde q}_+{\widetilde q}_- 
- \lambda_{23}^2\left({\widetilde s}_+{\widetilde q}_-
+ {\widetilde s}_+{\widetilde q}_-\right)^2
\end{eqnarray}

{}The effective low energy couplings of 
the type $\lambda_{22}\,{\bf Q\,Q\,{\widetilde Q}{\widetilde Q}}$ and 
$\lambda_{23} {\bf Q\,S\,{\widetilde Q}{\widetilde S}}$ 
which are responsible for 
generation--antigeneration pairing can be 
summarized in the following $2\times 5$ matrix:
\begin{center}
\begin{equation}
\label{16-16}
\begin{tabular}{l|l|l|c}
\multicolumn{2}{l}{$\hspace{30mm}{\bf\widetilde Q}_+$}&
\multicolumn{2}{l}{$\hspace{21mm}{\bf\widetilde Q}_-$} 
\\[1mm]
\cline{2-3}&&&\\[-2mm]
 ${\bf Q}_0$ & $\hspace{21mm}0$ & $\hspace{21mm}0$&\\[1mm]
 ${\bf Q}_{++}$ 
& $\hspace{2mm}\lambda_{22}\,q_{--}{\widetilde q}_- + 
\lambda_{23}\,s_{--} {\widetilde s}_-\hspace{2mm}$
& $\hspace{2mm}\lambda_{22}\,q_{--}{\widetilde q}_+ + 
\lambda_{23}\,s_{--} {\widetilde s}_+\hspace{2mm}$
&\\[1mm]
 ${\bf Q}_{--}$
& $\hspace{2mm}\lambda_{22}\,q_{++}{\widetilde q}_- + 
\lambda_{23}\,s_{++} {\widetilde s}_-$
& $\hspace{2mm}\lambda_{22}\,q_{++}{\widetilde q}_+ + 
\lambda_{23}\,s_{++} {\widetilde s}_+$
&\\[1mm]
 ${\bf Q}_{+-}$
& $\hspace{2mm}\lambda_{22}\,q_{-+}{\widetilde q}_- + 
y\lambda_{23}\,s_{-+} {\widetilde s}_-$
& $\hspace{2mm}\lambda_{22}\,q_{-+}{\widetilde q}_+ + 
\lambda_{23}\,s_{-+} {\widetilde s}_+$
&\\[1mm]
 ${\bf Q}_{-+}$
& $\hspace{2mm}\lambda_{22}\,q_{+-}{\widetilde q}_- + 
\lambda_{23}\,s_{+-} {\widetilde s}_-$
& $\hspace{2mm}\lambda_{22}\,q_{+-}{\widetilde q}_+ + 
\lambda_{23}\,s_{+-} {\widetilde s}_+$ 
&\\[3mm] 
\cline{2-3}
\end{tabular}
\end{equation}
\end{center}

When the fields ${\bf Q}_i, {\bf S}_i, {\bf\widetilde Q}_\pm$ and 
${\bf\widetilde S}_{\pm}$ acquire {\em vev}\,s 
(denoted in the table above as $q_i, s_i, {\widetilde q}_\pm$ and 
${\widetilde s}_{\pm}$,
respectively), a generation--antigeneration pairing takes place. 
Generically, only one scenario of pairing is allowed by F-flatness constraints:

$\bullet$  $\widetilde{q}_+ = \widetilde{s}_+ = 0$, 
$\widetilde{q}_- \neq 0$ 
and $\widetilde{s}_- \neq 0$ (interchange of indices 
$+\,\leftrightarrow\,-$ clearly provides an equivalent choice). 
This choice of {\em vev}\,s satisfies (\ref{f3},\ref{flat-2}) 
and one is left to choose 
$q_i$ and $s_i$ so that (\ref{f1},\ref{flat-1}) are satisfied.
Only one generation--antigeneration pairing occurs:
${\bf\widetilde Q}_+$ pairs 
with a linear combination of ${\bf Q}_q$ and ${\bf Q}_s$ where
\begin{eqnarray}
\label{Qs}
{\bf Q}_q = \frac{(q\cdot {\bf Q})}{\bf q}\equiv 
(q_{++}{\bf Q}_{--} + q_{--}{\bf Q}_{++} + 
 q_{+-}{\bf Q}_{-+} + q_{-+}{\bf Q}_{+-})/{\bf q}~;\\
{\bf Q}_s = \frac{(s\cdot {\bf Q})}{\bf s}\equiv 
(s_{++}{\bf Q}_{--} + s_{--}{\bf Q}_{++} + 
 s_{+-}{\bf Q}_{-+} + s_{-+}{\bf Q}_{+-})/{\bf s}~;\nonumber\\
 {\bf s}\equiv \sqrt{\sum_i s_i^2}~,\hspace{30mm}
 {\bf q}\equiv \sqrt{\sum_i q_i^2}\nonumber 
\end{eqnarray} 
Mass term has the form 
$\left(\lambda_{22}\,{ q}\,{\widetilde q}_-\,{\bf Q}_q + 
 \lambda_{23}\,{ s}\,{\widetilde s}_-\,{\bf Q}_s\right){\bf\widetilde Q}_+$.
  

To achieve second generation--antigeneration pairing one has to 
consider non-vanishing higher--point couplings that are present in the 
superpotential of this model, 
such as 8--point couplings 
$(\chi_{++}\chi_{--} + \chi_{+-}\chi_{-+})^2\,
{\widetilde \chi_+}^2\,{\widetilde \chi_-}^2$ and 
$\chi_0(\chi_{++}\chi_{--} + \chi_{+-}\chi_{-+})^2\,
{\widetilde \chi_+}\,{\widetilde \chi_-}\,{\widehat \Phi}$. 
These couplings will modify the F-flatness conditions and allow
small non-vanishing ${\widetilde s}_+$ and ${\widetilde q}_+$. 
As a result, the second generation--antigeneration pairing can take place at 
a lower scale.
(Here we note that F-flatness conditions will be modified once the 
supersymmetry is broken, so that the second generation--antigeneration
 pairing scale is very likely to be at least as high as $m_S$.)


{}Since the presence of light   
color Higgs triplets will mediate unacceptably rapid proton decay, 
one needs to arrange for them
to be appropriately heavy. One needs also to arrange for a pair of Higgs
doublets to remain light.
To simplify the analysis, we assume that Higgs doublets and triplets
come solely from {\bf 10} of $SO(10)$ (due to the mixing, they may also 
come from ${\bf 16}+\overline{\bf 16}$\,s). 

We give the adjoint $\Phi$ of $SO(10)$ a {\em vev} in the following 
direction: $\Phi=\epsilon
\otimes {\mbox{diag}}(a,a,a,b,b)$, where $a,b\neq 0,\;a\neq b$. 
Since ${\widetilde s}_+ = {\widetilde q}_+ = 0,\,\, 
{\widetilde h}_-,\,{\widetilde h}_-',\,{\widetilde H}_-$ and 
${\widetilde H}_-'$ decouple (they may become massive via the 
coupling ${\tilde\chi}^3_- {\tilde U}_+ D_+ D_- {\hat\Phi}^4$ at a 
much lower scale) and one is left with $6\times 6$ 
mass matrix ${\cal M}_{ij}$ of Higgs doublet couplings 
$(\lambda_{11}'\phi\,+\,\lambda_{11}^a\,b)s\,h\,h'$ and
$(\lambda_{23}'{\widetilde s}s+\lambda_{22}' 
{\widetilde q}q)({\widetilde h}\,h'+{\widetilde h}'\,h)$ 
(mixing is disregarded, the coupling constants 
are not shown explicitly in the table):

\begin{center}
\begin{equation}
\begin{tabular}{c|cccccc|c}
\multicolumn{2}{l}{$\hspace{10mm}h_0$} 
& $\hspace{8mm}h_{++} \hspace{8mm}$ & $\hspace{8mm}h_{--}\hspace{8mm}$
                   & $\hspace{8mm}h_{+-} \hspace{8mm}$ & 
$\hspace{8mm}h_{-+} \hspace{6mm}$ 
& \multicolumn{2}{l}{$\hspace{4mm}{\widetilde h}_+ $} 
\\[1mm]
\cline{2-7}&&&&&&&\\[-2mm]
& 0 & $\left(\phi + b\right) s_{--}$ &  $\left(\phi - b\right) s_{++}$
&  $\left(\phi + b\right) s_{-+}$ & $\left(\phi -b\right)s_{+-}$ 
   & 0 & $h_0' $\\[2mm]
& $\left(\phi - b\right) s_{--}$&0
& $\left(\phi + b\right) s_0 $ &0&0
   & $p_{--}$ & $h_{++}'$ \\[2mm]
& $\left(\phi + b\right) s_{++}$ &$\left(\phi - b\right) s_0 $&0&0&0
   & $p_{++}$ & $h_{--}'$ \\[2mm]
& $\left(\phi - b\right) s_{-+}$ & 0 & 0 & 0 & $\left(\phi + b\right) s_0 $ 
   & $p_{-+}$ & $h_{+-}'$ \\[2mm]
& $\left(\phi + b\right) s_{+-}$&0&0&$\left(\phi - b\right) 
s_0 $& 0 
   & $p_{+-}$ & $h_{-+}'$ \\[2mm]
&0&$p_{--}$ & $p_{++}$
 &$p_{-+}$ & $p_{+-}$ & 0
  & ${\widetilde h}_+' $\\[1mm] 
\cline{2-7}
\end{tabular}
\end{equation}
\end{center}
where we denote 
$p_i= \lambda_{23}'{\widetilde s}_-\,s_i 
+ \lambda_{22}'{\widetilde q}_-\,q_i$~.
The coupling constants $\lambda_i'$ are corresponding $\lambda_i$ from 
the (\ref{sp-so10}) corrected by Clebsch-Gordan factors upon gauge symmetry 
breaking (\ref{10-break}). 
Generically, ${\rm rank}({\cal M}_{ij}) = 6$. Thus, to have
${\rm rank}({\cal M}_{ij}) = 5$, {\em i.e.} one pair of 
light Higgs doublets, it is necessary to fine-tune the 
parameters entering the Higgs doublet matrix, {\em e.g.}, adjoint {\em vev} $b$.

{}Let us begin with the case $b=0$ 
(as in Dimopoulos--Wilczek mechanism). With some algebra, it can be shown that
$\det({\cal M}_{ij})$ vanishes only when 
$(s\cdot s)\,=\,(s\cdot q)\,=\,(q\cdot q)\,=0$ (this is the only choice that 
satisfies the F-flatness conditions (\ref{f1},\ref{f2})). However, 
with these conditions imposed on the {\em vev}\,s $s_i$ and $q_i$, 
the rank of ${\cal M}_{ij}$ invariably becomes four, {\em i.e.}, there
are {\em two} pairs of light Higgs doublets.
The extra pair of light Higgs doublets would affect
the successful SUSY GUT 
prediction for ${\mbox{sin}}^{2}\theta_{W}$. One thus has to consider 
fine--tuning $b$ to some non--zero value in order to arrange for only 
one pair of doublets to exist. Indeed, one finds that if $b$ is fine-tuned
to satisfy
\begin{eqnarray}
\label{b}
0&=&(p\cdot s)^2\,b^4\,+\,\phi^4\left((p\cdot s)^2
\,-\,(p\cdot p)(s\cdot s)\right)\\
&-&b^2\,\phi^2\,\left(2\,(p\cdot s)^2\,-\,
(p\cdot p)(s\cdot s)\,+\,8(p_{++}s_{+-}\,-\,p_{+-}s_{++})
(p_{--}s_{-+}\,-\,p_{-+}s_{--})\right)\nonumber
\end{eqnarray} 
then ${\rm rank}({\cal M}_{ij})\,=\,5$.   
 
{}Thus, there are three constraints imposed on the 13 parameters that enter 
the Higgs doublet mass matrix: $s_i$, $q_i$, $\widetilde{s}_-$,
$\widetilde{q}_-$, $s_0$, $\phi$ and $b$. 
These are the F-flatness conditions (\ref{f1},\ref{f2}) and condition 
(\ref{b}) for a pair of Higgs doublets to be light (constraints 
(\ref{f3},\ref{f4}) were already satisfied by setting 
${\widetilde s}_{+}={\widetilde q}_+=0$). The
D-flatness constraint (\ref{D-flat}) can be satisfied for arbitrary 
${\widetilde q}_-,\,q_i$ by tuning $q_0$ which otherwise does not enter our 
analysis. The resulting moduli space generically has only one pair of light 
Higgs doublets. An extra pair(s) of light Higgs doublets occur 
in the subspaces specified by $b=0$ or $s_0=0$.

The general expression for the light Higgs doublets is rather cumbersome 
and we choose to substitute constraint (\ref{b}) with two simpler constraints
from which (\ref{b}) follows directly:
\begin{eqnarray}
\label{bb}
0&=&p_{--}s_{-+}\,-\,p_{-+}s_{--}\;;\hspace{10mm}\phi-b=0
\end{eqnarray} 
In this subspace of the moduli space, the light Higgs doublet ${\widehat h}$ is
a particular linear combination of the full set of fields $h_0,\,h_i$ and 
${\widetilde h}_-$. Its dependence on $h_0$ will be of importance later on.   

The mass matrix of the Higgs triplet couplings ${\cal N}_{ij}$ has 
contributions from $(\lambda_{11}''\phi\,+\,\lambda_{11}^A\,a)s\,H\,H'$ and
$(\lambda_{23}''{\widetilde s}\,s~+~
\lambda_{22}''{\widetilde q}\,q)\left({\widetilde H}\,H'
\,+\,{\widetilde H}'\,H\right)$. It has a similar form as
${\cal M}_{ij}$ except $b$ is changed to $a$ and
the coupling constants are different. Generically, it has rank 6
and thus all Higgs triplets acquire a mass which is slightly below
$M_{GUT}$ (except ${\widetilde H}_-$ and ${\widetilde H}_-'$ as discussed
earlier).

So far we have specified
only the first generation--antigeneration pairing (\ref{Qs}). 
Let us denote that generation as ${\bf Q}_X$. The modified 
F-flatness conditions will allow for small non-vanishing 
${\widetilde q}_+$ and ${\widetilde s}_+$ and the second generation 
(which we denote as ${\bf Q}_Y$) will pair as 
can be seen from (\ref{16-16}). 
Let us for brevity write ${\bf Q}_X$ and ${\bf Q}_Y$ as 
\begin{eqnarray}
\label{QX-QY}
{\bf Q}_X&=&\frac{(l\cdot {\bf Q})}{\bf l}\equiv 
(l_{++}{\bf Q}_{--} + l_{--}{\bf Q}_{++} + 
 l_{+-}{\bf Q}_{-+} + l_{-+}{\bf Q}_{+-})/{\bf l}~;\hspace{16mm}
 {\bf l}\equiv \sqrt{\sum_i l_i^2}\\
{\bf Q}_Y&=&\frac{(k\cdot {\bf Q})}{\bf k}\equiv 
(k_{++}{\bf Q}_{--} + k_{--}{\bf Q}_{++} + 
 k_{+-}{\bf Q}_{-+} + k_{-+}{\bf Q}_{+-})/{\bf k}~;\hspace{10mm}
 {\bf k}\equiv \sqrt{\sum_i k_i^2}\nonumber
\end{eqnarray} 
where $l_i$ can be read from (\ref{Qs}) and $k_i$ have an analogous form with
${\widetilde q}_-,\,{\widetilde s}_-\,\rightarrow\,
{\widetilde q}_+,\,{\widetilde s}_+$. It is important to note that neither
${\bf Q}_X$ nor  ${\bf Q}_Y$ have a ${\bf Q}_0$ component.
Let us now analyze the mass matrix for the remaining three generations.
We have to choose a new basis 
${\bf Q}_0,\,{\bf Q}_1,\,{\bf Q}_2$ such that there are no mass terms of 
the form ${\bf Q}_{1,2}{\bf Q}_X$ or ${\bf Q}_{1,2}{\bf Q}_Y$. 
Without loss of generality we choose (the overall normalization factors 
are not 
shown)
\begin{eqnarray}
\label{Q12}
{\bf Q}_1&\sim& 
(l_{+-}k_{++}-k_{+-}l_{++}){\bf Q}_{--}+
(l_{--}k_{+-}-k_{--}l_{+-}){\bf Q}_{++}+
(l_{++}k_{--}-k_{++}l_{--}){\bf Q}_{+-}\\
{\bf Q}_2&\sim&
(l_{-+}k_{++}-k_{-+}l_{++}){\bf Q}_{--}+
(l_{--}k_{-+}-k_{--}l_{-+}){\bf Q}_{++}+
(l_{++}k_{--}-k_{++}l_{--}){\bf Q}_{-+}\nonumber
\end{eqnarray} 
The mass matrix for ${\bf Q}_0,\,{\bf Q}_1$ and ${\bf Q}_2$ is
\begin{center}
\begin{equation}
\begin{tabular}{c|cccc|}
\multicolumn{2}{l}{$\hspace{10mm}{\bf Q}_0$} 
&$\hspace{8mm}{\bf Q}_1\hspace{8mm}$ & 
\multicolumn{2}{l}{$\hspace{4mm}{\bf Q}_2$} \\[1mm]
\cline{2-5}&&&&\\[-2mm]
${\bf Q}_0$ & 0 &$a_{01}$ &$a_{02}$ &\\[2mm]
${\bf Q}_1$ &$a_{01}$ & $a_{11}$ & $a_{12}$&\\[2mm]
${\bf Q}_2$ &$a_{02}$ &$a_{12}$ & $a_{22}$ &\\[2mm]
\cline{2-5}
\end{tabular}
\end{equation}
\end{center}
The coefficients $a_{11},\,a_{12}$ and $a_{22}$ are proportional to $h_0$ and
since the light Higgs generically depends on $h_0$ they do not vanish.
The coefficients $a_{01}$ and $a_{02}$ depend on $h_i$ and are also generically
non-zero. Thus this matrix has rank three unless some fine-tuning is imposed
on the $s_i,\,q_i,\,{\widetilde s}_\pm$ and ${\widetilde q}_\pm$. It remains 
to be shown whether one can arrange for the rank of the quark mass matrix to 
be one as suggested by phenomenology. Should the answer be positive it will 
require fine-tuning of the {\em vev}\,s in some rather specific manner. 

Let us summarize the phenomenological properties of the $E_6$ and $SO(10)$
models analyzed in this preliminary analysis: \\
$\bullet$ ({\em i}\,) The mass of one generation--antigeneration pair
is $\approx 10^{16}$ GeV while that of the other pair is (perhaps, much) lower.
In our analysis, we assume that the generation and antigeneration come
entirely from ${\bf 16}$ and $\overline{\bf 16}$ of $SO(10)$.
{\em A priori}, after spontaneous symmetry breaking, the ${\bf 10}$ 
of $SU(5)$ in ${\bf 16}$ will mix with the ${\bf 10}$ of $SU(5)$ in the 
adjoint ${\bf 45}$ of $SO(10)$ (and similarly for the $\overline{\bf 16}$).
Clearly, a more careful analysis taking into account this mixing will
be useful. \\
$\bullet$ ({\em ii}) In a generic solution to the flatness conditions,
the Higgs triplet mass matrix has rank 6.
One can arrange for a pair of Higgs doublets to remain light
by fine-tuning the {\em vev} of the adjoint in its coupling to the doublets. 
We have chosen a scenario in which the Higgs fields come
solely from the ${\bf 10}s$ of $SO(10)$, that is, mixing with ${\bf 16}s$ 
has been ignored.
It is important to investigate 
possible mixing more carefully. \\ 
$\bullet$ ({\em iii}\,) Notice that only one generation-antigeneration
pairing takes place at $M_{GUT}$ 
and fine-tuning is necessary for the Higgs doublets to remain
light. There is no light Higgs doublet arising naturally in this model, nor 
could one obtain a quark mass matrix with rank one in a natural way.
These naturalness problems may result from the approximations we have made,
for instance, the K\"{a}hler potential, the higher order terms 
in the superpotential and non-perturbative 
corrections were ignored. 
Perhaps, stringy symmetries such
as target space modular invariance would improve our
understanding of the superpotential
and solve the fine-tuning problems in a natural way.\\
$\bullet$ ({\em iv}) The hidden sector gauge group $SU(2)$ is
asymptotically-free. Assuming that stabilization of the 
dilaton expectation value at the correct value results from the dynamical 
SUSY breaking, and evaluating the scale of the gaugino condensation to be 
$\sim\!\!10^{13}GeV$, we find that the $SUSY$ breaking scale in the 
visible sector is of order $10\sim 100~TeV$. \\
$\bullet$ ({\em v}) The $SU(3)_c$ and the $SU(2)_w$ adjoint Higgs fields are 
presumably quite light; they may be as light as the electroweak scale. 

Of course, all the above properties depend crucially on the non-perturbative 
string dynamics that we have assumed. Clearly, better understanding of 
SUSY breaking in string theory is needed. On the other hand, we see that
the spectra and their couplings are tightly restricted in the string models.
This already allows non-trivial tests of the viability of such models. 
It is clearly important to calulate explicitly the
perturbative couplings in the superpotential and the K\"{a}hler potential. 
This will provide further 
stringent tests on the viability of the $3$-family $E_6$ and its related 
$SO(10)$ string models. 

\acknowledgments

{}We thank Costas Bachas, Zurab Berezhiani, Keith Dienes, Lance Dixon, Gia Dvali, Alon Faraggi, Larry Hall, Michael Peskin, Joe Polchinski, Philippe Pouliot, Stuart Raby and especially Pran Nath for discussions. 
The research of G.S., S.-H.H.T. and Y. V.-K. was partially
supported by National Science Foundation. G.S. would also like to thank
Joyce M. Kuok Foundation for financial support. The work of Z.K. was
supported in part by the grant NSF PHY-96-02074, and the DOE 1994 OJI award.
Z.K. would also like to thank Mr. Albert Yu and Mrs. Ribena Yu for
financial support.

\end{document}